\documentclass[twoside,10pt]{article}
\usepackage{fleqn,espcrc2}
\usepackage{times,mathptm}
\usepackage[dvips]{graphicx}
\newcommand{\fcaption}{%
\vspace*{-1cm}
\caption%
}
\newcommand{\beq}{\begin{equation}}
\newcommand{\eeq}{\end{equation}}
\newcommand{\bed}{\begin{displaymath}}
\newcommand{\eed}{\end{displaymath}}
\newcommand{\bea}{\begin{eqnarray}}
\newcommand{\eea}{\end{eqnarray}}

\renewcommand{\b}{\beta}

\newcommand{\m}{\mu}
\newcommand{\s}{\sigma}

\renewcommand{\th}{\theta}

\newcommand{\dg}{\dagger}
\newcommand{\non}{\nonumber}

\newcommand{\AmS}{{\protect\the\textfont2
  A\kern-.1667em\lower.5ex\hbox{M}\kern-.125emS}}

\title{Vortex Structure in Abelian-Projected Lattice Gauge Theory}

\author{J. Ambj{\o}rn\address{The Niels Bohr Institute,
DK-2100 Copenhagen \O, Denmark}, J. Giedt \address{Physics Dept., 
Univ. of California, Berkeley, CA 94720 USA}, 
J. Greensite\address{Physics Dept., San Francisco State Univ., 
San Francisco, CA 94132 USA}\thanks{Talk presented by J. Greensite.  
Work supported by the U.S. Department of Energy under Grant 
No.\ DE-FG03-92ER40711.} }

\begin{document}
 
\begin{abstract}

  We report on a breakdown of both monopole dominance and positivity
in abelian-projected lattice Yang-Mills theory.  The breakdown is
associated with observables involving two units of the abelian charge.
We find that the projected lattice has at most a global $Z_2$ symmetry
in the confined phase, rather than the global U(1) symmetry that might
be expected in a dual superconductor or monopole Coulomb gas picture.
Implications for monopole and center vortex theories of confinement
are discussed.

\end{abstract}

\maketitle

   Center vortices can be located on thermalized lattices by
the technique of center projection in maximal center gauge, and their
effects on gauge-invariant observables such as Wilson loops and topological
charge have been clearly seen (e.g.\ in refs.\ \cite{Jan98},
\cite{dFE}, and in contributions to these Proceedings).  A competing
theory of confinement is the dual-superconductor/abelian-projection theory,
which has been intensively studied on abelian-projected lattices.
It is of some interest to ask if there is evidence of vortex structure
also on abelian-projected lattices and, if so, whether this structure is
consistent with a picture of the vacuum as a Coulomb gas of monopole loops
(for a more detailed presentation of this contribution, cf.\ \cite{j3}).

   There is already some evidence that center vortices, in the abelian
projection, would appear in the form of a monopole-antimonopole chain,
with the $\pm 2\pi$ monopole flux collimated (at fixed time) in tubelike
regions of $\pm \pi$ flux \cite{Zako}.  If this is so, then several qualitative
predictions follow, which can be tested numerically:
\begin{itemize}
\item There is $Z_2$, rather than $U(1)$, 
magnetic disorder on finite, abelian-projected lattices; 
\item  Monopole dominance breaks down for even
multiples of abelian charge; 
\item There is strong directionality of field strength around an
abelian monopole, in the direction of the vortex.
\end{itemize}
Consider large Wilson loops or Polyakov lines on the abelian-projected
lattice, corresponding to $q$ units of the abelian electric charge:
\bea
       W_q(C) &=& \Big< \exp[iq \oint dx^\m A_\m] \Big>
\non \\
       P_q &=& \Big< \exp[iq \int dt A_0] \Big>
\label{line}
\eea
Collimated $\pm \pi$ flux tubes cannot disorder $q=$ even Wilson loops
and Polyakov lines.  If these vortex tubes are the confining objects,
then only for $q=$ odd would we expect $P_q=0$, and an area law falloff
for Wilson loops.  In consequence, there would be $Z_2$, rather than
U(1), magnetic disorder/global symmetry in the confined phase.   

   In contrast, in the monopole Coulomb gas or dual superconductor
pictures, we would expect all multiples $q$ of electric charge to 
be confined; $P_q=0$ for all $q$.  This is inferred from saddlepoint 
calculations in $QED_3$
and strong-coupling calculations in $QED_4$, it appears to be true
for the dual abelian Higgs theory (a model of dual superconductivity),
as well as in a simplified treatment of the monopole gas in ref.\
\cite{HT}.  The $Z_2$ subgroup of U(1) plays no special role in the monopole
picture.
   
   Center vortices are rather thick objects $\sim 1$ fm, so at, e.g.,
$\b=2.5$ we would need $12\times 12$ $q=2$ Wilson loops to see string
breaking.  This is impractical.  The ``fat link'' technique is 
untrustworthy in this case, due to the absence of a transfer matrix,
and in any case rectangular loops (with $R \gg T$) are inadequate; the 
appropriate operator mixings have to be taken into account.  For these 
reasons, it is best to study $q=$ even Polyakov lines, rather than Wilson 
loops. 
\begin{figure}[t!]
\centerline{\includegraphics[width=1.0\linewidth]{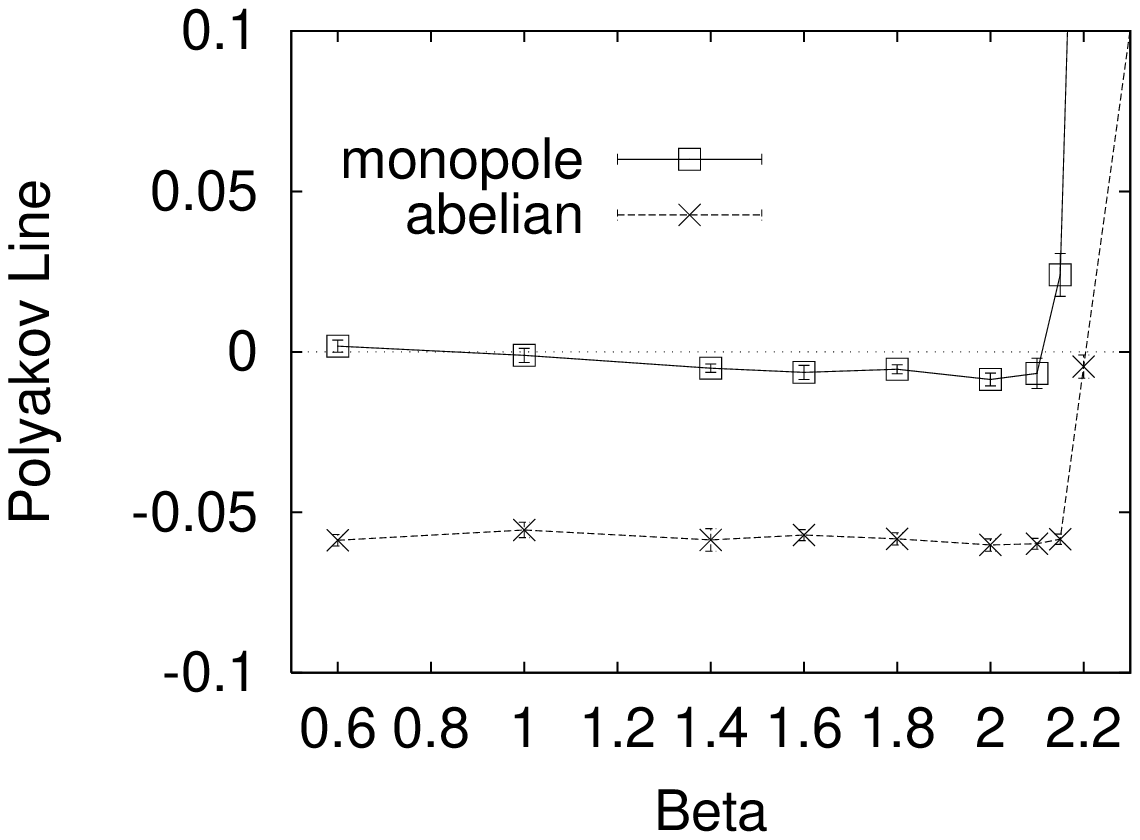}}
\fcaption{$q=2$ Polyakov lines, $T=3$.}
\label{P2T3a}
\end{figure}

  Figure \ref{P2T3a} shows our data, in the confined phase, for $q=2$
abelian Polyakov lines on a $12^3 \times 3$ lattice.  The upper line
is the monopole dominance (MD) approximation for this
quantity, following the method of \cite{Suzuki}.  We see that $P_2$
is finite, negative, and that there is a severe breakdown of the MD
approximation in this case.  The negative sign is allowed by
the absence of reflection positivity in maximal abelian gauge.
The finiteness of $P_2$ is expected in the center vortex picture,
and implies $Z_2$ rather than U(1) disorder on the abelian lattice,
while the breakdown of the MD approximation indicates that the
abelian monopole flux is \emph{not} distributed Coulombically.

  It is possible to avoid positivity problems by fixing to a
spacelike maximal abelian gauge \cite{Poli}
\beq
 R = \sum_x \sum_{k=1}^3 \mbox{Tr} [\s_3 U_k(x) \s_3 U^\dg_k(x)]
    ~~~ \mbox{is maximized}
\eeq
What happens in this case is that the loss of positivity is 
replaced by a breaking of $90^\circ$ rotation symmetry.  Spacelike
$P_2$ lines remain negative.  Timelike $P_2$ lines become positive,
although much smaller in magnitude, on a hypercubic lattice, than the 
spacelike lines.  Since this is a physical gauge, the result means that 
$q=2$ electric charge is unconfined.

   It is also interesting to write the link phase angles 
$\th_\m(x)$ of the abelian link variables as a sum 
of the link phase angles $\th^M_\m(x)$ in the MD approximation, 
plus a so-called ``photon'' contribution 
$\th^{ph}_\m(x) \equiv \th_\m(x) - \th_\m^M(x)$.
It is known that the photon field has no confinement properties at 
all \cite{Suzuki}; Polyakov lines constructed from links 
$U_\mu=\exp[i\th^{ph}_\m]$  are finite (also at higher $q$), 
and corresponding Wilson loops have no string tension.  Since $\th_\m^M$ 
would appear to carry all the confining properties, a natural conclusion 
is that the abelian lattice is indeed a monopole Coulomb gas.

   To see that this conclusion may be mistaken, suppose we 
\emph{add}, rather than subtract, the MD angles to the abelian
angles, i.e.
\beq
       \th'_\m(x) = \th_\m(x) + \th_\m^M(x)
                  = \th^{ph}_\m(x) + 2\th_\m^M(x)
\label{add}
\eeq
in effect doubling the strength of the monopole Coulomb field.
In the monopole picture, this doubling would be expected to increase the
$q=1$ string tension, with $P_1$ remaining zero.  Surprisingly, the 
opposite occurs; we in fact find that $P_1$ is negative in the
$\th'$ configurations, with values shown in Table 1.
\begin{table}[h!]
\centerline{
\begin{tabular}{|l|r|r|} \hline\hline
  T       & $\b$  &  $P_1$ line \\ \hline
  3       &  1.8  &  -0.0299 (20) \\ 
  3       &  2.1  &  -0.0405 (10) \\
  4       &  2.1  &  -0.0134 (10) \\ \hline
\end{tabular} }
\caption{$q=1$ Polyakov lines on the $\th'$ lattice.}
\label{table1}
\end{table}
  What this indicates is that the ``photon'' and MD contributions
do \emph{not} factorize in Polyakov lines and Wilson loops, contrary
to the case in the Villain model.  In fact, there is an important and
non-perturbative correlation between Polyakov line phases $\th^{ph}$ 
and $\th^M$, with the former breaking the (near) U(1) symmetry of the MD lattice down to an
exact $Z_2$  symmetry.  For example, if one computes the average value
of $\th^{ph}$ for $\th^M$ in the intervals $[0,{\pi\over 2}]$ and
$[{\pi\over 2},\pi]$ ($\b=2.1,T=3$), one finds
\beq
       \overline{\th}^{ph} = \left\{ \begin{array}{rl}
          0.027(4) & \mbox{for~~} \th^M \in [0,{\pi \over 2}] \cr
         -0.027(4) & \mbox{for~~} \th^M \in [{\pi \over 2},\pi]
         \end{array} \right. 
\label{oline}
\eeq
The question, of course, is what is the origin
of this correlation.  From the standpoint of the vortex theory,
what is happening is that the Coulombic distribution of $2\pi$ monopole
flux in the MD approximation is modified, by its correlation with
$\th^{ph}$, into a configuration with an exact $Z_2$ remnant symmetry;
confining flux has the same magnitude on the abelian
projected and MD lattices, but is distributed differently (collimated
vs.\ Coulombic) at large scales.  The negative value of $P_1$ in the additive
$\th'$ configurations can actually be deduced from the negative value of 
$P_2$ on the abelian projected lattice.  For this, we refer the interested 
reader to ref.\ \cite{j3}.

\begin{figure}[t!]
\centerline{\includegraphics[width=1.0\linewidth]{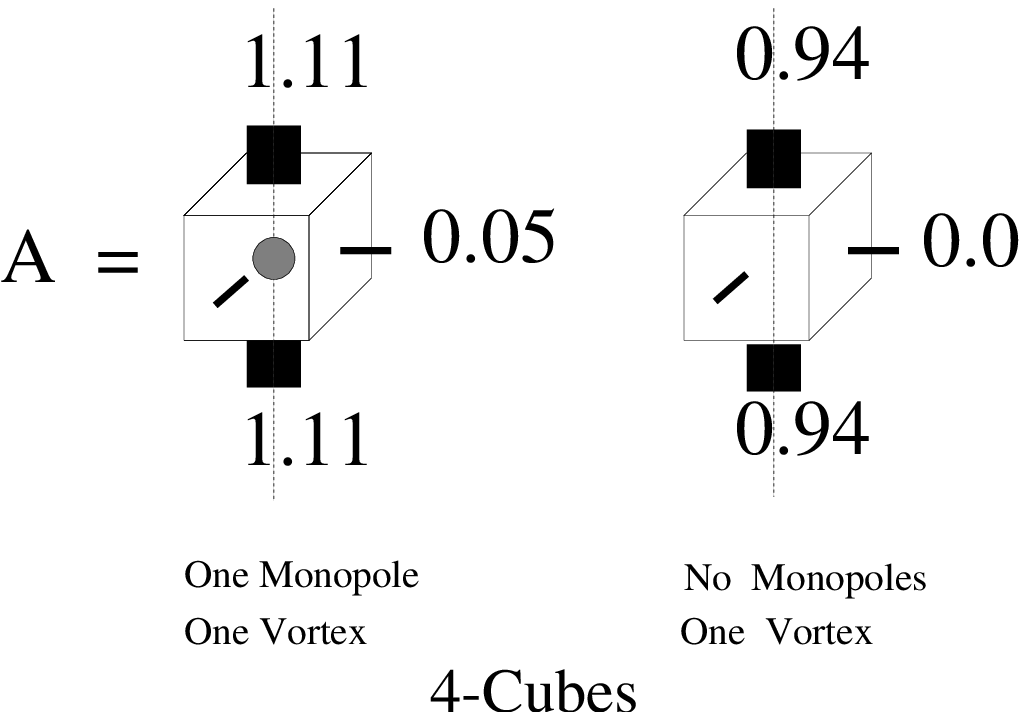}}
\fcaption{Field asymmetry around one and zero monopole 4-cubes.
The dashed line indicates cube faces pierced by a P-vortex.}
\label{mcube4}
\end{figure}

   Finally, one would like to see the collimation of field strength, in
the neighborhood of an abelian monopole, more directly. Here we have
extended the original efforts in ref.\ \cite{Zako} in two ways:  First,
in the indirect maximal center gauge, we have verified that there is an
almost exact alternation of monopoles with antimonopoles along P-vortex
lines, as previously conjectured.  In the few exceptional cases, there
is a static monopole or antimonopole within one lattice spacing of the 
P-vortex which, if counted as lying along the P-vortex, would restore the 
exact alternation.   Secondly, we have considered spacelike cubes 
$N=1-4$ lattice spacings wide, pierced on two faces by a single P-vortex,
and containing either one or zero static abelian monopoles.  We define  
$W_n^M(N,N)$ as the vev of unprojected Wilson loops, bounding 
faces of an $N\times N$ cube containing one static monopole.  The subscript
$n=0,1$ indicates that the face is pierced ($n=1$) or unpierced ($n=0$)
by a P-vortex line. $W_n^0(N,N)$  is the corresponding data for spacelike
cubes containing no monopole currents.  We then define the fractional
deviations
\bea
  A^M_{0,1} = {W^0_0(N,N) - W^M_{0,1}(N,N) \over W^0_0(N,N)}
\non \\
  A^0_{0,1} = {W^0_0(N,N) - W^0_{0,1}(N,N) \over W^0_0(N,N)}
\eea
The result for 4-cubes is shown in Fig.\ \ref{mcube4}. It is clear that the
flux is correlated very strongly with the P-vortex direction, and only
rather weakly with the presence or absence of a monopole inside the
cube.  This is what is expected in the center vortex picture.
   
   We conclude that the (i) non-confinement of $q=$ even abelian electric
charge; (ii) breakdown of the monopole dominance approximation; and 
(iii) highly asymmetric distribution of confining fields around monoples,
is consistent with vortex structure on the abelian lattice, but is
probably not compatible with monopole Coulomb gas or dual superconductor
pictures.  An important point is that charged fields (e.g.\ off-diagonal
gluons) in a confining theory, even if very massive, can have a profound 
effect on infrared structure. We think it likely that the monopole 
Coulomb gas and dual-superconductor pictures also break down
in the D=3 Georgi Glashow and the Seiberg-Witten models, respectively
(cf.\ the discussion in refs.\ \cite{j3,j2}), albeit on a $q=2$ 
string-breaking scale which increases with the mass of the W-bosons.


\begin{thebibliography}{9}
\bibitem{Jan98} L. Del Debbio, M. Faber, J. Giedt, J. Greensite, 
\v{S}. Olejn{\'\i}k, Phys. Rev. D58(1998) 094501, hep-lat/9801027.
\bibitem{dFE} Ph. de Forcrand and M. D'Elia, Phys. Rev. Lett. 82 (1999)
4582, hep-lat/9901020.
\bibitem{j3} J. Ambj{\o}rn, J. Giedt, J. Greensite, hep-lat/9907021.
\bibitem{Zako} L. Del Debbio, M. Faber, J. Greensite, 
\v{S}. Olejn{\'\i}k, Zakopane proceedings, hep-lat/9708023.
\bibitem{HT} A. Hart and M. Teper, hep-lat/9902031.
\bibitem{Suzuki} T. Suzuki et al., Phys. Lett. B347 (1995) 375,
hep-lat/9408003.
\bibitem{Poli} M. Chernodub, M. Polikarpov, A. Veselov, JETP Lett. 69
(1999) 174, hep-lat/9812012.
\bibitem{j2} J. Ambj{\o}rn and J. Greensite, JHEP (1998) 9805:004,
hep-lat/9804022.
\end{thebibliography}
\end{document}